# HIGH VOLTAGE ELECTROSTATIC PENDULUM

**Raju Baddi**

**National Center for Radio Astrophysics, TIFR, Ganeshkhind P.O Bag 3, Pune University Campus, PUNE 411007, Maharashtra, INDIA; baddi@ncra.tifr.res.in**

**ABSTRACT**

A pendulum powered by high voltage electricity is described. The pendulum consists of two conducting plates(thin foil) separated by copper rods and are insulated from each other. High voltage is applied to these plates through the connecting copper rods. Another stationary aluminum plate(thin foil) is placed in front of the pendulum such that it serves to attract the pendulum plates and makes electrical contact with them enabling charge transfer between the stationary plate and the pendulum plates. The pendulum is powered by the energy stored in the capacitance between the stationary aluminum plate and the pendulum plate. Attempt has been made to obtain the time period of oscillations as a function of applied voltage and other parameters. The derived formula for the time period has been verified experimentally. This apparatus can be used to demonstrate electrical phenomena in general and in particular electrical energy stored in conductors of small dimensions.

## 1. INTRODUCTION

A conductor connecting to a source of constant potential difference like the terminal of a battery achieves the same potential as the terminal of the source with respect to the ground terminal. This causes charges to flow between the conductor and the terminal. The amount of charge being decided by the conductor's size, shape and its position with respect to the ground terminal. Technically this is accounted by the capacitance of the conductor in question. With this we say the capacitor(here the conductor making contact with the terminal of the source in the presence of other grounded conductors) is charged to the potential difference of the source. The capacitance of connecting wires in an electrical circuit have a very low value, typically ~pf or sub pf. The energy stored in a capacitor is given by $½CV^2$ (Griffith's 2000). The energy is directly proportional to the capacitance while is so to the square of the voltage. The energy store in low valued capacitor like that formed by conductors of small dimension can be increased by having a large voltage. For example a 0.1pF capacitor charged to a voltage of 5V has 1.25pJ of energy. However the same capacitor when charged to a voltage of 1000V would have $4x10^4$ times more energy. This energy can be considered for work on a small mechanical device, like a pendulum. For instance an energy of $½CV^2$=50nJ stored in the capacitor when converted into mechanical/kinetic energy $½mv^2$ of an object of 100mg would imply a velocity of ~3cm/s. In this view the frictional energy loss(here air drag) in the movement of the pendulum given its total mass is of the order quoted above can be compensated by the electrical energy stored in the capacitance of a conductor and hence sustaining its oscillations indefinitely under proper conditions. This energy store determines the rate at which the pendulum can dissipate energy. This in turn determines the period of the pendulum. This manuscript is divided into six sections. The first section gives an introduction to the problem, the construction of the electrostatic pendulum is described in the second section. Third section discusses the energy aspects. Time period of oscillations are derived in fourth section. Section five presents the experimental verification of the relations derived in section 4. Sixth section forms the conclusion.

## 2. THE ELECTROSTATIC PENDULUM

The electrostatic pendulum consists of two metallic conducting plates(Al foil ~2cm x 3cm and 0.1mm thick) separated by connecting rods(copper wire ~0.5cm in diameter and 4 cm long). These rods do not make any electrical contact with each other and are used to suspend the pendulum vertically using two small rings as shown in Figure 1. Further another stationary conducting plate which is sufficiently broad to cover both the plates is placed in front of the pendulum. This plate serves to attract the pendulum plates providing the necessary acceleration to the plates. Through the connecting rods high voltage(>400V DC) is applied to the plates. The pendulum is free to oscillate like the conventional pendulum. However being light in weight and having broad plates it is unlikely to exhibit oscillations under gravitational acceleration due to easy damping. The oscillations are mainly governed by electrostatic attraction between the pendulum plate and the stationary plate. In this view the gravitational acceleration is not taken into consideration in spite of vertical mount.

Figure 2 shows the various stages of oscillation and how the stationary plate is charged in different phases. The oscillatory behavior of the pendulum is understood as follows. As a charged metallic conductor can attract another uncharged conductor we start from the moment when the upper plate(also called positive plate), which is incidentally charged positively, makes contact with the stationary plate. This makes the stationary plate to come to the same potential as the upper plate w.r.t the lower plate(also called grounded plate), which we assign to be at zero potential or simply grounded. Now the stationary plate is charged positively such that capacitance it forms with the grounded plate develops a potential difference equal to that between the pendulum plates or in other words the high voltage supply. Once the potential of the stationary plate is the same as that of the positive plate there will be no electric field to cause attraction between them. On the other hand now the stationary plate attracts the grounded plate. This makes the pendulum to shift to its second phase which is the motion of grounded plate towards the stationary plate. When the grounded plate makes contact with the stationary plate it discharges it and brings it to the ground potential. Now that the stationary plate has achieved the same potential as the lower plate the attraction between them seizes. But the attraction between the upper plate and the stationary plate takes over. The whole process is repeated indefinitely. The pendulum can be easily constructed from commonly available materials like aluminum foil, commercial copper wire (0.5mm dia) and high voltage plastic for structural requirements. However the required high voltage source has to be constructed using appropriate electronic circuit(Millman & Halkais and references there in).

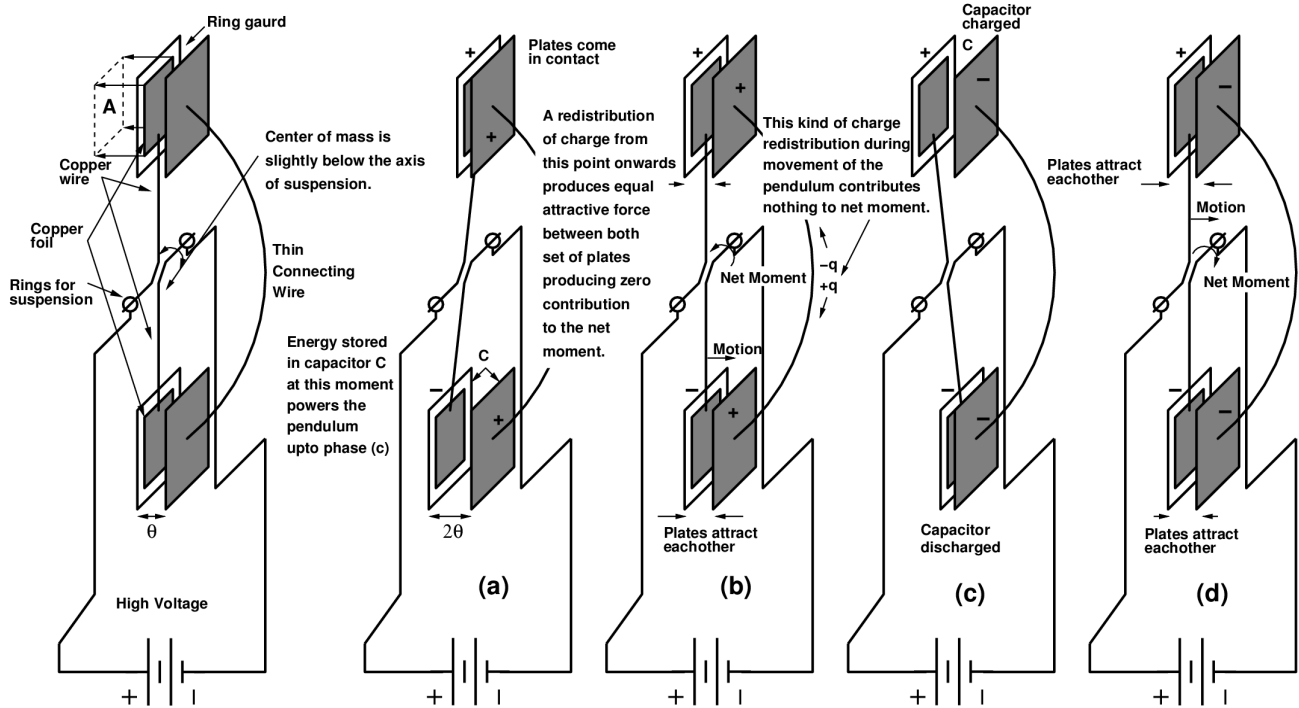

*Fig 1 : Left : Construction of the pendulum with center of gravity slightly below the axis of suspension so that the pendulum takes a vertical stand. High voltage could be any where >400V. Right (a,b,c,d): Various phases of oscillations, the length of the arrow indicates a measure of force.*

Since the power required to sustain oscillations of the pendulum is small, a low power high voltage generator is sufficient. The attractive force depends on the potential difference between the pendulum plates and the stationary plate. In fact it is directly proportional to this voltage. Doubling the potential difference would double the attractive force at every point during the motion of the pendulum producing twice the acceleration. Further a charge transfer occurs during each contact of pendulum plates with the stationary plate. This is essentially a discrete current flow between the pendulum plates via the stationary plate which serves as a temporary reservoir of charge. The next section briefly discusses upon the quantitative aspects of electrical/mechanical energy in the pendulum.

## 3. ENERGY CONSIDERATIONS

The stationary plate and the pendulum plates form a capacitor. Due to symmetry this capacitor has the same value C for both the plates as shown in Figure 2. The energy stored in this capacitor is converted into mechanical energy by mutual attraction between the stationary plate and the pendulum plate. As the pendulum plate moves towards the stationary plate the energy stored in the capacitor reduces while the kinetic energy of the plates increases. This has been shown by giving the analog of a simple parallel plate capacitor to the left of Figure 2. The lower plate is grounded while the upper plate is charged with a constant charge Q, such that when the upper plate of the pendulum touches the stationary plate charge Q flows from the pendulum plate to the stationary plate and increases its potential to +V. Once this happens the attractive force between the upper plate and the stationary plate vanishes while the lower plate is attracted towards the stationary plate. The energy stored in the capacitance between the pendulum plates and the stationary plate( C, right of Figure 2) is now converted into mechanical energy of the pendulum. Considering the simplified case of parallel plate capacitor, we see that the energy E stored in a capacitor $C_P$ when the plates are separated by a distance *d* is given by (Griffith's 2000)

$$E = \frac{1}{2}C_P V^2 = \frac{1}{2}\frac{\epsilon_0 A}{d}V^2 \quad (1)$$

where A is the area of the plates and V is the potential difference between them. Potential difference can also be written in terms of the charge Q and capacitance C, so we have

$$E_0 = \frac{1}{2}\frac{\epsilon_0 A}{d_0}\left(\frac{Q}{C_0}\right)^2 = \frac{1}{2}\frac{d_0}{\epsilon_0 A}Q^2;$$
$$E = \frac{1}{2}\frac{\epsilon_0 A}{d}\left(\frac{Q}{C}\right)^2 = \frac{1}{2}\frac{d}{\epsilon_0 A}Q^2; \quad (2)$$

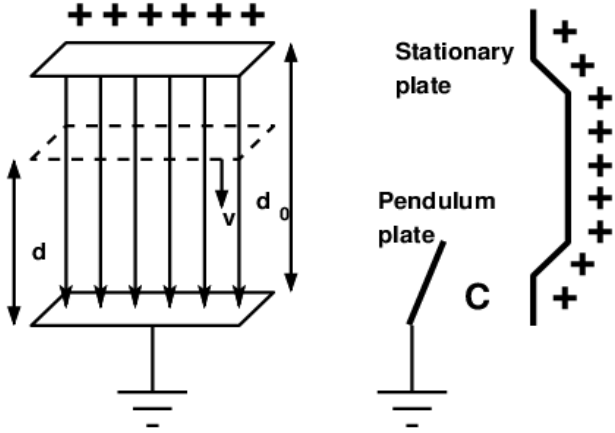

*Fig 2 : Energy stored in the capacitance between the stationary plate and the pendulum plates is converted into mechanical energy when the pendulum plates move towards the stationary plate by attraction. Left : Analog of a parallel plate capacitor. The initial distance between the plates is $d_0$, when the distance between the plates reduces to **d** a part of capacitor's energy is converted into kinetic energy of the plates. Right: Actual configuration of the capacitor C in the case of pendulum that stores the energy. Since the configuration is complicated to calculate the capacitance certain assumptions simplify the situation. Which eventually leads to conversion of energy in C, ½CV $^2$ to mechanical energy ½mv$^2$ of the plates. Thus providing the required drag force that overcomes air resistance and sustains the pendulum's oscillations.*

Since the field between the capacitor plates is always constant the force on the moving plate is also constant. Such that the work done $E_M$ on it to increase its mechanical energy is $F_E(d_0 - d)$. Where $F_E$ is the force on the upper plate due to the grounded lower plate. Using the method of image charges(Griffith's 2000) the grounded lower plate can be replaced by another plate exactly similar to the upper plate but bearing negative charge -Q instead of positive +Q and placed below the grounded plate at the same distance as the positively charged plate above it. Thus we see the force $F_E$ is given by

$$F_E = \frac{1}{2}\frac{Q^2}{\epsilon_0 A} \quad ; \quad E_M = \frac{1}{2}\frac{d_0 - d}{\epsilon_0 A}Q^2 \quad (3)$$

We see from (3) and (2) $E+E_M=E_0$. When the upper plate meets the grounded plate all the energy stored in the capacitor is converted into mechanical energy. In the case of pendulum we assume that the pendulum plate and the region of the stationary plate in the vicinity of the pendulum plate form a much larger capacitance than the rest of the stationary plate. So for the duration of close encounter we still have a parallel plate capacitor configuration and the situation is similar to that of the analog. With this we assume that the net mechanical energy transferred to the pendulum is ½$CV^2$. C can be either calculated approximately by assuming parallel plate configuration between the plates or measured. The next section derives an expression for the period of the pendulum by considering the dissipation of the energy gain of the pendulum by air drag force on the moving plates.

## 4. TIME PERIOD OF OSCILLATIONS

The time period of oscillations is calculated by assuming the pendulum plates to be moving at a constant speed. This results in a constant air drag force on the plates which resists the motion of plates through air(Strelkov 1987). Further the collision of pendulum plates with the stationary plate is assumed to be perfectly elastic and takes place within a negligible time compared to the period of oscillation. The plates are assumed to instantly bounce back once they come in contact with the stationary plate. With the plates moving at a constant speed *v* and the separation distance between the pendulum plates in equilibrium position and the stationary plate being θ we can write for the time period T as

$$T = \frac{4\theta}{v} \qquad (4)$$

Now since the plates are assumed to be moving in air at a constant speed *v* we write for the drag force $F_{drag}$ (Strelkov 1987) as

$$F_{drag} = \frac{\rho v^2 A}{2} C_x \qquad (5)$$

where ρ is the density of air and $C_x$ is the drag coefficient which in turn depends on the Reynolds number $R_e$ which is a function of velocity of air flow(here *v*). The work done by the drag force during half a period is $F_{drag}$ x 2θ. This is compensated by the energy stored in the capacitance C between the pendulum plates and the stationary plate as discussed in section 4. We equate work done by (5) on both the plates to ½$CV^2$ and obtain the speed *v* of the plates,

$$2\theta \rho v^2 A = \frac{1}{2}CV^2 \;;\; v = \frac{V}{2}\sqrt{\frac{C}{\theta \rho A}} \qquad (6)$$

Here we have chosen for the sake of simplicity $C_x$=1.0. From (4) we have the time period as,

$$T = \frac{4\theta}{v} = \frac{8\theta}{V}\sqrt{\frac{\theta \rho A}{C}} \qquad (7)$$

Approximate calculations with a prototype produce values of the same order for observed time period. However the inverse proportionality of T with V has been tested experimentally and has been found to hold good. This has been discussed in section 5.

## 5. EXPERIMENTAL VERFICATION

Experiments with a prototype indicate that the inverse relation of time period T with applied high voltage V in (7) is valid. Also approximate calculations produce values of T of the same order as observed for voltage range of 600-900V even though it was not possible to verify this thoroughly. The capacitance C being in sub-picofarad range was difficult to measure, only an approximate calculation assuming a parallel plate configuration could be used. Some of the difficulties with the prototype were variation of separation distance θ along the length of the pendulum plates due to inclined orientation, possible contribution from pivot friction, incomplete overlap between plates, asymmetry in orientation of plates and frictional issues of contact between pendulum plates and the stationary plate especially at low frequencies(~1Hz) which sometimes stopped the oscillations. Table 1 gives a comparison between calculated and observed values. Even though a good agreement between calculated and observed values is seen, this agreement is not claimed strongly due to uncertainty in the value of capacitance C(~0.16pF) and applied voltage V.

| Applied voltage V (volts) | $T_{observed}$ seconds | $T_{calculated}$ seconds |
|---|---|---|
| 4$V_{ac}$ | 0.62 | 0.61 |
| 5$V_{ac}$ | 0.52 | 0.50 |
| 6$V_{ac}$ | 0.43 | 0.41 |

*Table 1: Comparison of calculated values as per (7) with experimental values.*

## 6. CONCLUSION

This article describes an electrostatic pendulum that can be easily constructed and used to demonstrate conversion of electrical energy into mechanical energy. We also see that the frequency of the oscillations is directly proportional to the applied voltage V (7). This suggests another use for the measurement of voltage. This would be possible by replacing the pendulum plates by conducting elastic leaves on a high voltage insulating base. The frequency of the vibration of the leaves would be proportional to the potential difference between them. The vibration frequency can be obtained by having an optical/audio coupler.